\documentclass[aps,prl,twocolumn,superscriptaddress,nobibnotes]{revtex4-1}
\usepackage{graphicx}
\usepackage{color}
\usepackage{hyperref}
\usepackage{amssymb}
\usepackage{amsmath}
\usepackage{float}
\usepackage{tabularx}
\usepackage{nicefrac}

\begin{document}

\title{Comparing fermionic superfluids in two and three dimensions}

\author{Lennart Sobirey}
\email{lsobirey@physnet.uni-hamburg.de}
\affiliation{Institut f\"{u}r Laserphysik, Universit\"{a}t Hamburg}

\author{Hauke Biss}
\affiliation{Institut f\"{u}r Laserphysik, Universit\"{a}t Hamburg}
\affiliation{The Hamburg Centre for Ultrafast Imaging, Universit\"{a}t Hamburg, Luruper Chaussee 149, 22761 Hamburg}

\author{Niclas Luick}
\affiliation{Institut f\"{u}r Laserphysik, Universit\"{a}t Hamburg}
\affiliation{The Hamburg Centre for Ultrafast Imaging, Universit\"{a}t Hamburg, Luruper Chaussee 149, 22761 Hamburg}

\author{Markus Bohlen}
\affiliation{Institut f\"{u}r Laserphysik, Universit\"{a}t Hamburg}
\affiliation{The Hamburg Centre for Ultrafast Imaging, Universit\"{a}t Hamburg, Luruper Chaussee 149, 22761 Hamburg}

\author{Henning Moritz}
\affiliation{Institut f\"{u}r Laserphysik, Universit\"{a}t Hamburg}
\affiliation{The Hamburg Centre for Ultrafast Imaging, Universit\"{a}t Hamburg, Luruper Chaussee 149, 22761 Hamburg}

\author{Thomas Lompe}
\affiliation{Institut f\"{u}r Laserphysik, Universit\"{a}t Hamburg}
\affiliation{The Hamburg Centre for Ultrafast Imaging, Universit\"{a}t Hamburg, Luruper Chaussee 149, 22761 Hamburg}

\date{\today}

\maketitle
\textbf{
Understanding the origins of unconventional superconductivity has been a major focus of condensed matter physics for many decades.
While many questions remain unanswered, experiments have found that the systems with the highest critical temperatures tend to be layered materials where superconductivity occurs in two-dimensional (2D) structures \cite{Keimer2015-xc,Ge2015-es,Yu2019-py}.
However, to what extent the remarkable stability of these strongly correlated 2D superfluids is related to their reduced dimensionality is still an open question \cite{Yu2019-py}.
In this work, we use dilute gases of ultracold fermionic atoms \cite{Bloch2008-td,Ketterle2008-ca} as a model system to directly observe the influence of dimensionality on strongly interacting fermionic superfluids.
We achieve this by measuring the superfluid gap of a strongly correlated quasi-2D Fermi gas \cite{Levinsen2015-lf,Turlapov2017-wg} over a wide range of interaction strengths and comparing the results to recent measurements in 3D Fermi gases \cite{Biss2021-ir}.
We find that the superfluid gap follows the same universal function of the interaction strength in both systems, which suggests that there is no inherent difference in the stability of fermionic superfluidity between two- and three-dimensional quantum gases.
Finally, we compare our data to results from solid state systems and find a similar relation between the interaction strength and the gap for a wide range of two- and three-dimensional superconductors. 
}

Fermionic particles such as the electrons in superconductors have half-integer spin and therefore obey the Pauli exclusion principle.
This prevents systems of noninteracting fermions from condensing into a macroscopic wavefunction and becoming superfluid.
However, in the presence of an effective attractive interaction it can become energetically favorable for fermions with opposite spin to form bosonic pairs.
These pairs can then condense into a coherent many-body state and become superfluid, as laid out by Bardeen, Cooper and Schrieffer (BCS) in their famous theory of superconductivity \cite{Bardeen1957-ti}. 
The energy that is required to break one of these pairs is called the superfluid gap $\Delta$, as the pairing manifests itself as a gap in the excitation spectrum of fermionic superfluids.
Since breaking the pairs destroys the superfluid, the size of this gap determines the stability of the superfluid and sets its critical temperature. 

Over the last decades, new classes of superconductors have been discovered that exhibit higher critical temperatures and stronger interactions than conventional BCS superconductors \cite{Uemura1991-gy}. 
Of particular interest are systems where superfluidity occurs in two-dimensional structures, as they are the ones where the highest ambient-pressure critical temperatures have been observed \cite{Keimer2015-xc}.  
However, the dimensionality of these systems cannot be changed without dramatically altering their other properties as well, and it is therefore unclear to what extent the surprising stability of their superfluidity is predicated on their two-dimensional nature.

In this work, we directly observe the effect of reduced dimensionality on the stability of strongly interacting fermionic superfluids. 
We measure the superfluid gap of an ultracold 2D Fermi gas as a function of interaction strength and compare the results with our recent measurement of the gap in a three-dimensional system \cite{Biss2021-ir}.
We find that the superfluid gap follows the same universal function of the chemical potential in both systems, which suggests that dimensionality has only limited influence on the stability of strongly interacting fermionic superfluids.

\begin{figure*}[htbp]
\begin{center}
\includegraphics[width=18.3cm]{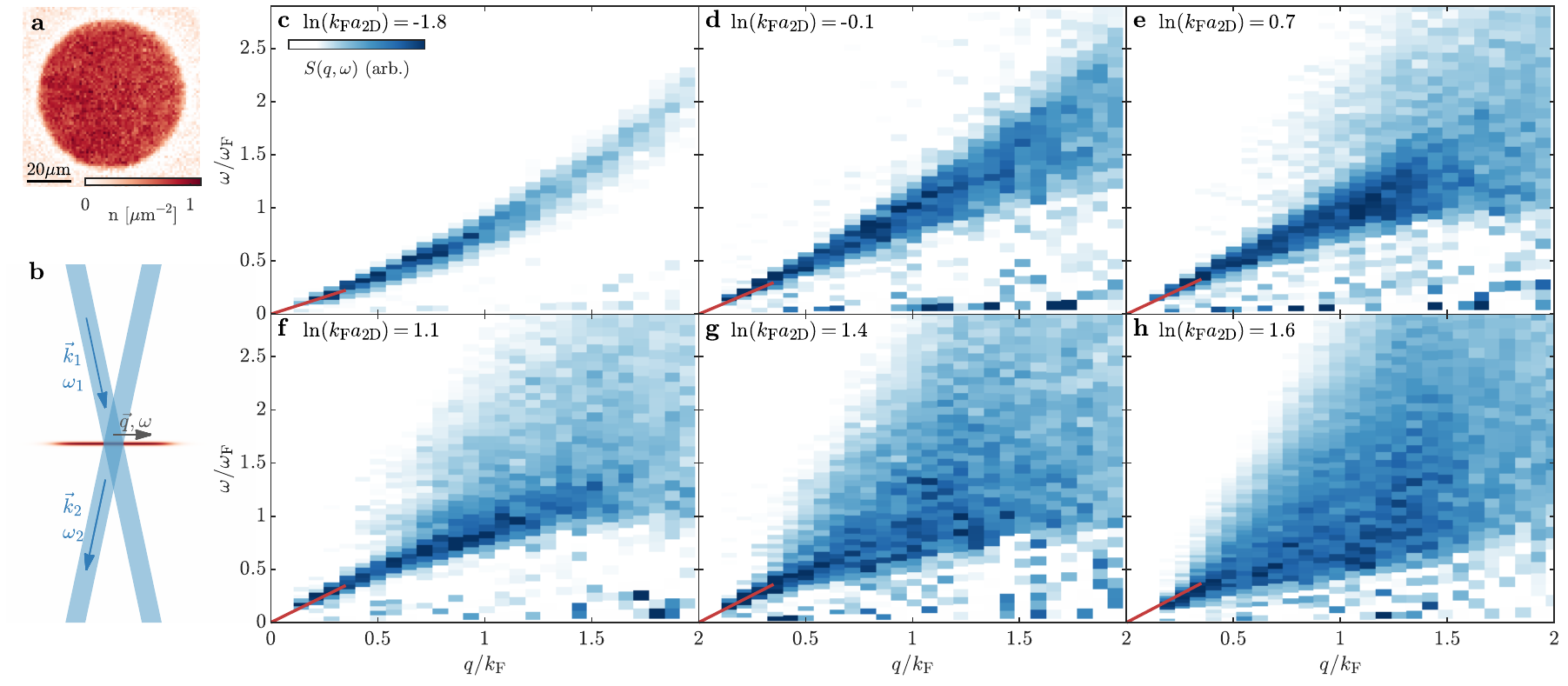}
\caption{\textbf{Excitation spectrum of a 2D Fermi gas in the BEC-BCS crossover.}
\textbf{a}: Absorption image showing the density $n(\vec{r})$ of our homogeneous 2D Fermi gas. 
\textbf{b}: Sketch of the experimental setup for measuring the excitation spectrum of our system.
Two far-detuned laser beams drive a two-photon transition with energy and momentum transfer $\hbar \omega = \hbar (\omega_1 - \omega_2)$ and $\hbar q = \hbar |\vec{k}_1-\vec{k}_2|$, and the dynamic structure factor $S(q,\omega)$ can be obtained from the resulting heating rate.
\textbf{c-h}: Measurements of $S(q,\omega)$ taken at different values of the 2D interaction parameter $\rm{ln}(k_{\rm F}a_{\rm 2D})$.
For strong attractive interactions (\textbf{c}), the system consists of tightly bound molecules which are excited as unbroken pairs, and consequently $S(q,\omega)$ shows the Bogoliubov dispersion of an interacting Bose gas.
Moving into the crossover regime (\textbf{d,e,f}), the pairs become more weakly bound and pair breaking excitations begin to appear at higher momenta.
These excitations become more pronounced as we approach the BCS limit where the system shows the expected broad pair breaking continuum (\textbf{g,h}). 
In addition to these pair breaking excitations, it is also possible to excite sound waves in the superfluid.
These appear in our spectra as a linear mode at low momenta, with a slope that corresponds to the speed of sound in the system and is in excellent agreement with previous measurements (red lines in panels \textbf{c}-\textbf{h}, \cite{Bohlen2020-iu}).
The data in panel \textbf{a} (\textbf{c}-\textbf{h}) was obtained by averaging over 26 (3-8) individual measurements.
The behavior observed in (\textbf{c}-\textbf{h}) closely resembles results obtained in 3D Fermi gases \cite{Biss2021-ir} (for a comparison see Supplementary Materials). 
}
\label{FigIntro}
\end{center}
\end{figure*}

For our experiments, we use ultracold atomic Fermi gases of $^6$Li atoms.
Such gases have two key advantages that make them uniquely suited for performing experiments that isolate the effect of dimensionality on the stability of superfluids: 
The first is that they are systems with simple and well-understood interparticle interactions that can be easily tuned using Feshbach resonances \cite{Chin2010-th}.
The second is that the dimensionality of the system can be controlled freely by changing the shape of the confining potential \cite{Martiyanov2010-bx,Frohlich2011-qe,Sommer2012-dt,Ries2015-fe,Cheng2016-nd,Mitra2016-mo,Peppler2018-wg,Hueck2018-sx}.
By combining these two features, we can create two- and three-dimensional systems that have the same microscopic physics but different dimensionality.

To perform a quantitative comparison between these systems, we examine the effect of the reduced dimensionality on the superfluid gap.
The gap is well suited for this purpose, as it directly determines both the critical current and the critical temperature of a fermionic superfluid and thus constitutes an excellent measure for its stability.
As reliable measurements of the gap are available for three-dimensional Fermi gases \cite{Schirotzek2008-ig,Hoinka2017-ej,Biss2021-ir}, we can focus our experiments on measuring the gap in two-dimensional systems.

To bring our system into the two-dimensional regime, we apply a strong confining potential along one direction such that the chemical potential and temperature are well below the level spacing.
This strongly suppresses all excitations in this direction and thereby creates an effective (or quasi-) 2D system \cite{Petrov2001-qx,Levinsen2015-lf,Turlapov2017-wg} (see Supplementary Materials). 

We then measure the excitation spectrum of the gas to determine the size of the superfluid gap. 
We use momentum resolved Bragg spectroscopy to measure the dynamic structure factor $S(q,\omega)$ of the superfluid, which describes the probability of creating an excitation in the system by providing an energy and momentum transfer of $\hbar \omega$ and $\hbar q$ (Fig.~\ref{FigIntro}\textbf{b}, see Supplementary Materials).
By tuning the strength of the interparticle interactions, we can perform such measurements throughout the crossover from a BCS superfluid of weakly bound Cooper pairs to a Bose-Einstein-condensate (BEC) of deeply bound molecules, where the interaction strength can be parametrized by the 2D interaction parameter $\rm{ln}(k_{\rm F}a_{\rm 2D})$.
Here, $a_{\rm 2D}$ is the 2D scattering length \cite{Petrov2001-qx}, $k_{\rm F} = \sqrt{2 m E_{\rm F}}/\hbar$ is the Fermi wavevector, $E_{\rm F} = \hbar \omega_{\rm F}$ is the Fermi energy, and $m$ is the mass of a $^6$Li atom.

The results of these measurements are shown in Fig.~\ref{FigIntro}\textbf{c}-\textbf{h}.
We observe two different types of excitations \footnote{While in principle a third type of excitations exists in the form of the Higgs mode, this mode is expected to lie within the pair breaking continuum \cite{Zhao2020-tw}) and is therefore not visible in our spectra.}:
The first are longer-range collective excitations of the superfluid, which are visible as a linear sound mode at low momentum transfers ($q \ll k_{\rm F}$), with a slope that is in excellent agreement with the speed of sound measured in \cite{Bohlen2020-iu} (red lines in Fig.\,\ref{FigIntro}\,\textbf{c}-\textbf{h}).
This is the Goldstone mode of the system, which arises from the breaking of the U(1) symmetry of the system when the gas condenses into a superfluid \cite{Goldstone1961-hs,Hoinka2017-ej}.
The second type of excitations are single-particle excitations that break a pair.
As this process is only possible if the energy transfer is sufficiently high to overcome the energy gained from pairing, these excitations show a sharp onset at an energy transfer of $2 \Delta$.
This behavior is most apparent for BCS superfluids with weak attractive interactions (Fig.\,\ref{FigIntro}\,\textbf{g},\textbf{h}), where a pronounced continuum of pair breaking excitations is clearly visible.
When increasing the interparticle attraction, the size of the superfluid gap increases and consequently the onset of the pair breaking continuum shifts towards higher energies.
Additionally, as the pairs are transformed from weakly bound Cooper pairs to tightly bound bosonic molecules, the onset of the pair breaking continuum moves towards higher momenta as pair breaking excitations are suppressed when the size of the pairs becomes small compared to the length scale of the perturbation \cite{Leggett1998-bd,Buchler2004-hh}.
This trend continues into the BEC regime, where the molecules are so tightly bound that pair breaking excitations become completely suppressed.
The excitation spectrum then exhibits the well-known Bogoliubov dispersion relation of a superfluid Bose gas, which consists of a linear dispersion of phonons at low momentum and single-particle excitations of bosonic molecules at higher momenta (see Fig.\,\ref{FigIntro}\,c).  

\begin{figure}
\begin{center}
\includegraphics[width=8.9cm]{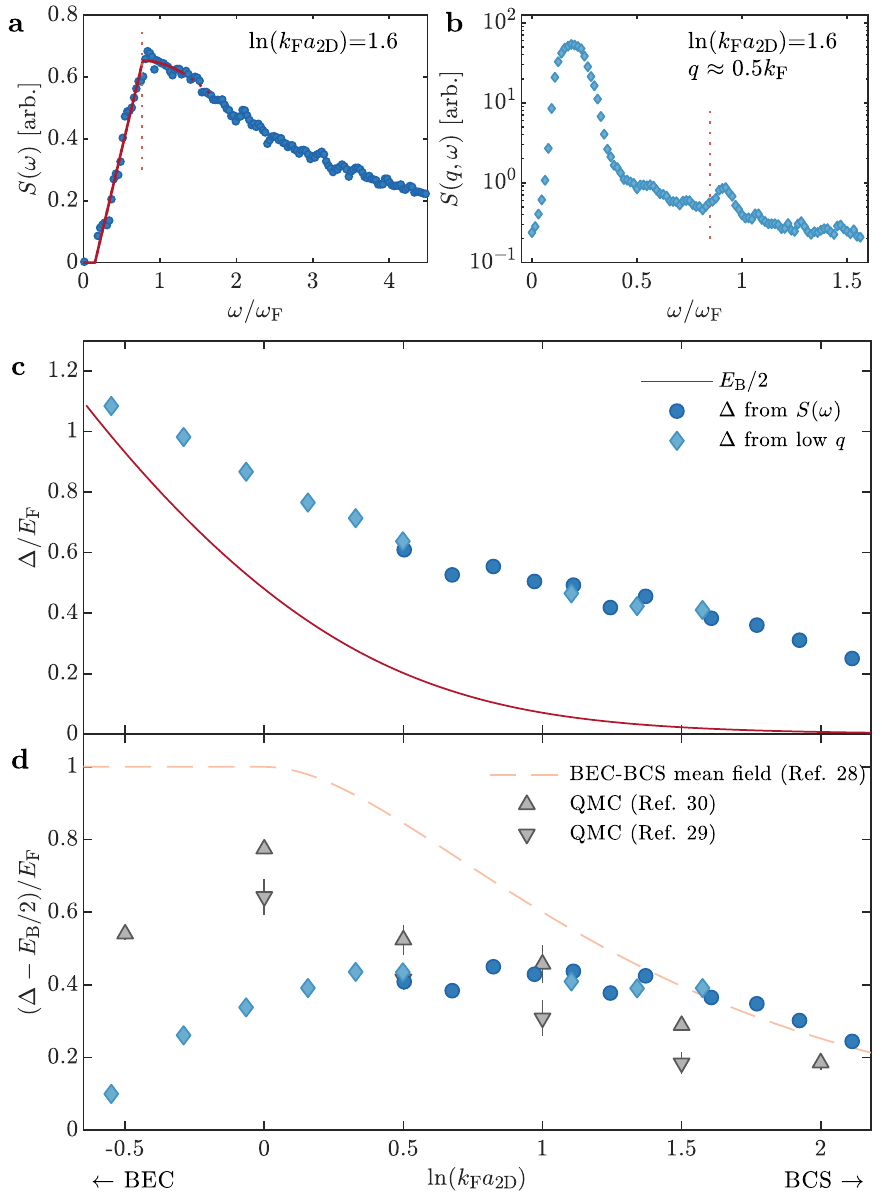}
\caption{\textbf{Superfluid gap in the 2D BEC-BCS crossover.}
\textbf{a}: Integrated dynamic structure factor $S(\omega) = \int S(q,\omega) q\,dq$ at an interaction strength of $\rm{ln}(k_{\rm F}a_{\rm 2D})=1.6$. 
We determine the energy $2\Delta$ (red dotted line) at which pair breaking excitations become possible from a phenomenological fit (solid red line) to $S(\omega)$ (see Supplementary Materials).
\textbf{b}: Dynamic structure factor $S(q,\omega)$ measured at a fixed momentum transfer of $q=0.5\,k_{\rm F}$.
At these small wavevectors, pair breaking is suppressed and strong driving is required to observe the onset of the pair breaking mode at $2\Delta$ (red dotted line), which causes a strong saturation of the low-energy phononic mode.
\textbf{c}: Measured superfluid gap $\Delta$ as a function of the interaction strength.
The different symbols distinguish results obtained using the approaches shown in panel \textbf{a} ($S(\omega)$, blue dots) and \textbf{b} (low $q$, light blue diamonds). 
The contribution of the two-body bound state to the gap is shown as a dark red line. 
\textbf{d}: Many-body contribution $\Delta - E_{\rm B}/2$ to the superfluid gap $\Delta$. 
BEC-BCS mean field predictions (dashed orange line, \cite{Randeria1989-qp}) are in agreement with our measurement in the BCS regime, but deviate in the strongly correlated crossover regime.
Quantum Monte-Carlo calculations (gray triangles, \cite{Vitali2017-hr,Zielinski2020-pa}) show better agreement in the crossover, but still deviate from the measurements in the BEC regime.
Error bars denote $1\sigma$ confidence intervals of the fit and are smaller than the symbol size, the data shown in panels \textbf{a} (\textbf{b}) is the average of 8 (26) individual measurements.
} 
\label{FigGap}
\end{center}
\end{figure}

To determine the superfluid gap $\Delta$ from our measurements, we integrate the measured dynamic structure factors over the momentum axis.
The resulting quantity $S(\omega) = \int S(q,\omega) q\,dq$ describes the probability of creating an excitation with a given energy $\hbar \omega$ and is similar to the Raman response measured in inelastic Raman scattering experiments in solid state physics \cite{Devereaux2007-ho}.
On the BCS side of the crossover, our measurements of $S(\omega)$ show the same behavior as observed in the Raman response of s-wave BCS superconductors: A sharp increase at $2\Delta$, followed by a slow decay back to zero \cite{Klein1984-yx} (Fig.~\ref{FigGap}\textbf{a}).
In the BCS regime, we can therefore directly extract the size of the superfluid gap from $S(\omega)$ (see Supplementary Materials). 

Towards the BEC regime, extracting quantitative information from $S(\omega)$ becomes more difficult as the onset of the pair breaking continuum is masked by an increasing weight of the Goldstone mode.
Fortunately, we can circumvent this problem by probing the system at lower momenta, where the energy of phononic excitations is significantly below $2\Delta$ \cite{Hoinka2017-ej,Biss2021-ir}.
While pair breaking excitations are suppressed at these low wavevectors, they can still be observed when using a stronger drive. 
This is shown in Fig.~\ref{FigGap}\textbf{b}, where the onset of the pair breaking mode at $2\Delta$ is clearly separated from the strongly driven Goldstone mode visible at lower energy.
Employing both of these methods enables us to measure the superfluid gap $\Delta$ throughout the BEC-BCS crossover. 
The resulting values are plotted as a function of the 2D interaction parameter $\rm{ln}(k_{\rm F}a_{\rm 2D})$ in Fig.~\ref{FigGap}\textbf{c}, together with the binding energy $E_{\rm B}$ of the bare two-body bound state (red line), which in 2D systems exists for any non-zero attractive interaction \cite{Levinsen2015-lf}.
For our smaller attractive interactions ($\rm{ln}(k_{\rm F}a_{\rm 2D}) \gtrsim 1.5$), the two-body binding energy is negligible, and the sizable gap of $\Delta \approx 0.3\,E_{\rm F}$ is entirely due to many-body effects.
However, when going into the crossover regime, the trivial two-body binding energy increases and becomes comparable to the effect of the many-body BCS pairing.
To separate these two contributions to the gap and thereby determine the evolution of the many-body contribution throughout the crossover, we subtract the known value of the two-body binding energy \cite{Levinsen2015-lf} from our measured gaps.
As can be seen in Fig.~\ref{FigGap}\textbf{d}, the many-body contribution $\Delta - E_{\rm B}/2$ grows with increasing interactions in the BCS regime, reaches a maximum in the crossover regime and then decreases again towards the BEC side of the resonance, where the contribution of the two-body bound state begins to dominate as the gas turns into a BEC of deeply bound molecules.
When comparing these results to theory, we find that they are in excellent agreement with mean-field theory \cite{Randeria1989-qp} in the BCS regime, but begin to deviate from the mean-field results in the strongly correlated crossover region ($\rm{ln}(k_{\rm F}a_{\rm 2D}) \approx 1$).
Quantum Monte-Carlo (QMC) simulations \cite{Vitali2017-hr,Zielinski2020-pa} are in somewhat better agreement with our data in the crossover, but still predict larger values of $\Delta - E_{\rm B}/2$ in the BEC regime.

\begin{figure}
\begin{center}
\includegraphics[width=8.9cm]{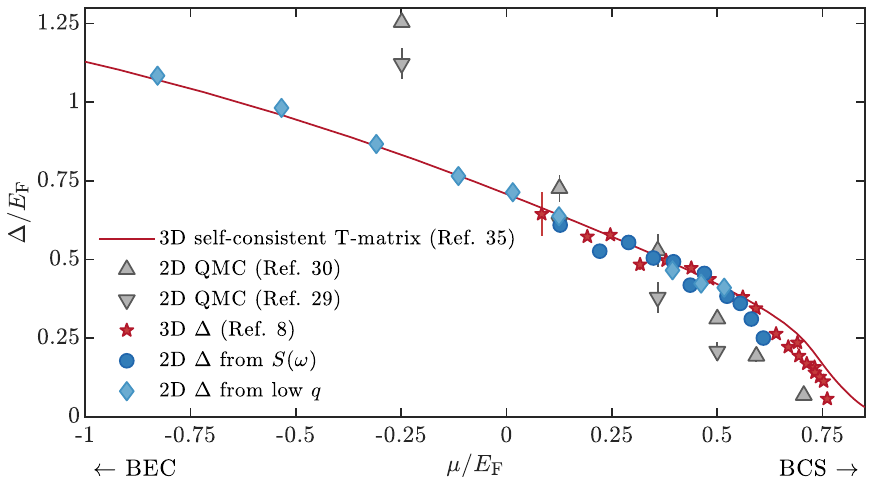}
\caption{\textbf{Comparing the gap of two-dimensional and three-dimensional superfluids.}
Superfluid gap $\Delta/E_{\rm F}$ of two-dimensional (blue circles and diamonds) and three-dimensional (red stars) Fermi gases as a function of the chemical potential $\mu/E_{\rm F}$, for which we use the results of QMC calculations \cite{Astrakharchik2004-qs,Shi2015-pd}.
The measurements of the gap collapse onto a single curve, which is well-described by theoretical predictions for the gap in three-dimensional Fermi gases (red line \cite{Haussmann2007-ph,NoteMu}); results from 2D QMC calculations \cite{Vitali2017-hr,Zielinski2020-pa} are shown as gray diamonds (see Supplementary Materials).
Error bars denote $1\sigma$ confidence intervals of the fit.
}
\label{FigComp}
\end{center}
\end{figure}

\begin{figure*}
\begin{center}
\includegraphics[width=18.3cm]{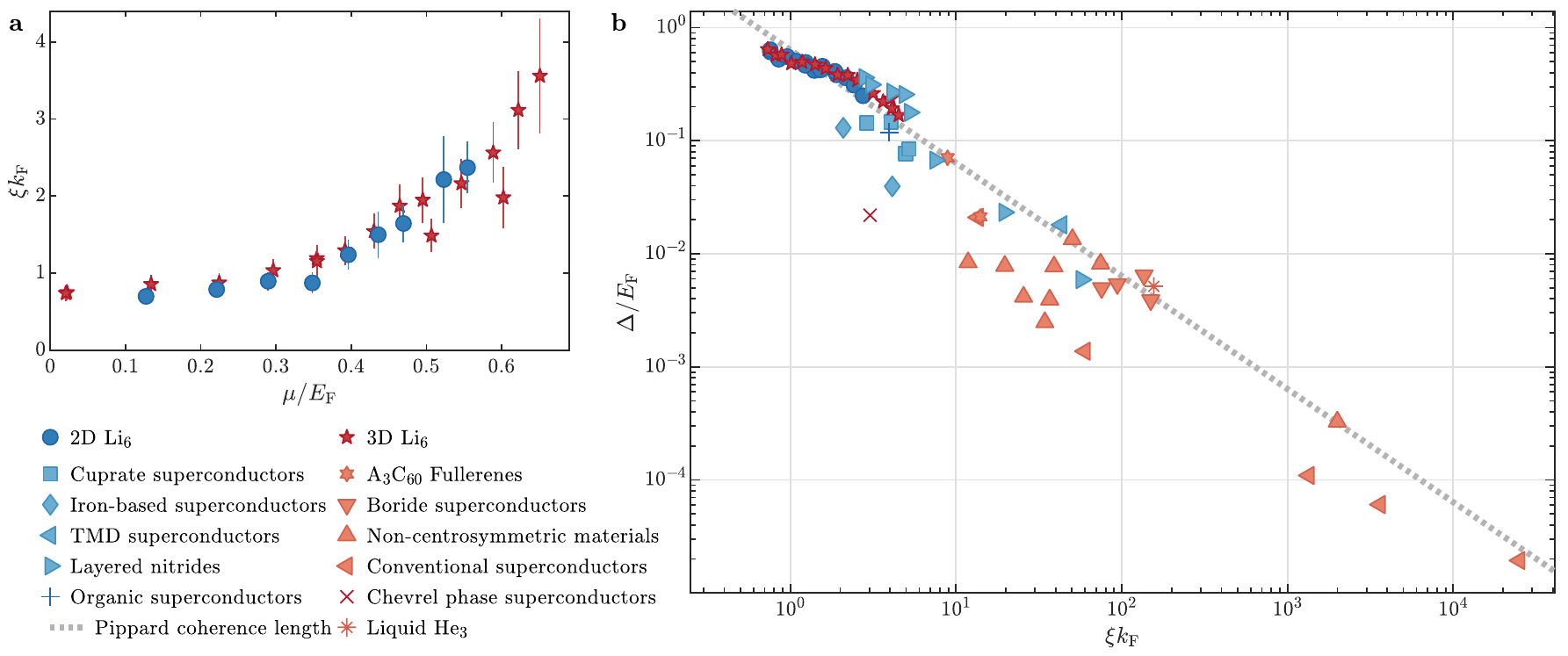}
\caption{\textbf{Fermionic superfluidity in different materials.}
\textbf{a}: Dimensionless pair size $\xi k_{\rm F}$ plotted as a function of the dimensionless chemical potential $\mu/E_{\rm F}$.
As $\xi k_{\rm F}$ follows the same function of $\mu/E_{\rm F}$ for the two- and three-dimensional systems, $\xi k_{\rm F}$ can be used as an alternative parametrization of the interaction strength in our strongly interacting Fermi gases.
Error bars denote $1\sigma$ confidence intervals. 
\textbf{b}: Plot of the dimensionless gap $\Delta/E_{\rm F}$ against the dimensionless pair size $\xi k_{\rm F}$ for different fermionic superfluids.
The Pippard coherence length $\xi_p = \frac{\hbar^2 k_{\rm F}}{\pi m \Delta}$ \cite{Ashcroft1976-up} is shown as a dotted line.
Remarkably, the superfluid gap is roughly proportional to the inverse of the dimensionless pair size for a wide range of materials which span five orders of magnitude in $\Delta/E_{\rm F}$ and $\xi k_{\rm F}$.
This observation applies equally to two- and three-dimensional systems, which suggests that strong correlations are more important for the stability of fermionic superfluids than the dimensionality of the system. 
}
\label{FigMat}
\end{center}
\end{figure*}

We now proceed to compare our measurements to recent results from 3D Fermi gases.
To perform such a comparison, we need to find a suitable parametrization of the interaction strength, as the dimensionless interaction parameters $\rm{ln}(k_{\rm F}a_{\rm 2D})$ and $1/k_{\rm F}a_{\rm 3D}$ that are commonly used in two- and three-dimensional systems parametrize the interactions differently and cannot be compared directly.
Instead, we parametrize the interaction strength with the normalized chemical potential $\mu/E_{\rm F}$ of the fermions.
This choice is motivated by the fact that the chemical potential is a basic thermodynamic quantity that is defined independent of dimensionality and has monotonous and well-known relations to the 2D and 3D interaction parameters $\rm{ln}(k_{\rm F}a_{\rm 2D})$ and $1/k_{\rm F}a_{\rm 3D}$ \cite{Shi2015-pd,Boettcher2016-hu,Astrakharchik2004-qs,Ku2012-wd}.
Therefore, we can perform our comparison by plotting the superfluid gap $\Delta/E_{\rm F}$ as a function of the chemical potential $\mu/E_{\rm F}$ for two- and three-dimensional systems. 
The results are shown in Fig.~\ref{FigComp}.

Remarkably, we find that within the accuracy of our measurements, the results for $\Delta/E_{\rm F}$ obtained in two- and three-dimensional systems collapse onto a single curve.
This suggests that for strongly interacting Fermi gases, the gap follows a single, universal function $f(\mu/E_{\rm F})=\Delta/E_{\rm F}$ of the interaction strength that is independent of the dimensionality of the system. 
The data appears to be well-described by theoretical predictions for three-dimensional fermionic superfluids \cite{Haussmann2007-ph}, but as shown in Fig.~\ref{FigGap}\textbf{d} deviates from 2D QMC calculations \cite{Vitali2017-hr,Zielinski2020-pa}.
This is unlikely to be the result of imperfect two-dimensional confinement or thermal excitations, which would both be expected to have the strongest effect in the BCS regime (see Supplementary Materials).
Consequently, our measurements imply that for a given coupling strength, there is no inherent difference in the stability of fermionic superfluidity between two- and three-dimensional quantum gases.

As we perform our experiments in an ideal model system, it is natural to ask to what extent our results apply to other, more complex materials.
The first step to answering this question is to extend Fig.\,\ref{FigComp} by adding data from other fermionic superfluids.
However, while the chemical potential provides an excellent measure for the interaction strength in our strongly interacting quantum gases, there is a large number of materials where it is not known with sufficient accuracy to be used as a parametrization of the interaction strength.
We therefore need a different interaction parameter that is viable in a wide variety of systems including ultracold gases, liquid $^3$He and solid state superconductors.
One parameter that has been suggested for this purpose is the dimensionless pair size $\xi k_{\rm F}$ \cite{Randeria1989-qp,Pistolesi1994-oe,Stintzing1997-av,Schunck2008-us,Marsiglio2015-qh}. 
Similar to the chemical potential, $\xi k_{\rm F}$ describes a fundamental physical property whose definition is independent of the dimensionality of the system.
Additionally, $\xi$ is closely related to the coherence length \cite{Pistolesi1996-ja}, which has been measured in many solid state superconductors. 

To obtain an estimate of the pair size for our systems, we consider the momentum dependence of the pair breaking excitations observed in our measurements of the dynamic structure factor in two- and three-dimensional Fermi gases \cite{Biss2021-ir}.
As discussed above, these measurements show a suppression of pair breaking excitations when the wavelength of the probing lattice becomes comparable to the size of the pairs. 
We can therefore fit the onset of the pair breaking continuum on the momentum axis and relate the onset momentum $\hbar k_o$ to the pair size via $k_o \sim 1/\xi$, with a prefactor that can be determined by comparing the results to theoretical predictions for two-dimensional \cite{Randeria1990-ox} and three-dimensional \cite{Marini1998-bb} systems (see Supplementary Materials).
We plot the resulting values of $\xi k_{\rm F}$ as a function of $\mu/E_{\rm F}$ and find that the results for the two- and three-dimensional systems collapse onto a single curve (Fig.~\ref{FigMat}\textbf{a}). 
This shows that our estimate of $\xi k_{\rm F}$ can be used as an alternate parametrization of the interaction strength in ultracold Fermi gases \cite{Marsiglio2015-qh}. 

Consequently, we can now plot our measurements of the gap $\Delta/E_{\rm F}$ as a function of the pair size $\xi k_{\rm F}$ and compare the results to a wide variety of different superconductors. 
The results are shown in Fig.~\ref{FigMat}\textbf{b}. 
Remarkably, all materials fall into a single band, which extends from conventional superconductors with gaps on the order of $10^{-5}\,E_{\rm F}$ and large coherence lengths to ultracold Fermi gases with gaps comparable to the Fermi energy and coherence lengths approaching the interparticle spacing, with a wide variety of exotic superconductors in between.
Fig.~\ref{FigMat}\textbf{b} therefore clearly shows a direct correlation between shorter coherence lengths and larger gaps \cite{Uemura1991-gy,Pistolesi1994-oe} that holds from the weak coupling limit all the way into the strongly correlated regime.
This correlation exists independent of the dimensionality of the material, in excellent agreement with our observations in two- and three-dimensional Fermi gases.
Therefore, our findings suggest that there is no inherent increase in the stability of a fermionic superfluid in two dimensions compared to a three-dimensional system with the same coupling strength.

In this work, we have used measurements of the excitation spectrum of strongly interacting ultracold Fermi gases to determine the superfluid gap and found that the gap follows a universal function of the interaction strength that is independent of the dimensionality. 
By extending the comparison to other fermionic superfluids, we have shown that this observation appears to hold for a wide range of two- and three-dimensional systems. 
Consequently, our results suggest that there is no inherent increase in the stability of the superfluid phase in lower dimensions.

Our work highlights that ultracold gases and strongly correlated superconductors can be realized at comparable effective interaction strengths.
In particular, the range of interaction strengths accessible with ultracold Fermi gases has significant overlap with the tuning range of the coupling strength of recently realized two-dimensional materials such as magic-angle twisted trilayer graphene \cite{Park2021-ug} and lithium-intercalated layered nitrides \cite{Nakagawa2021-ih}.
This raises the prospect of preparing ultracold gases and solid state systems that have the same effective interaction strength and directly comparing their properties.
For example, this could enable comparative studies of the transition from a superfluid to a strongly correlated normal state, which in 2D quantum gases has been observed to also show strong many-body pairing \cite{Murthy2018-en} and is difficult to study in many solid state systems due to the presence of competing order parameters \cite{Kondo2011-vz,Kondo2013-vq,Hashimoto2014-tm,Keimer2015-xc}.  

We thank G. Salomon, J. P. Brantut and L. Mathey for helpful comments on the manuscript.
This work is supported by the Deutsche Forschungsgemeinschaft (DFG, German Research Foundation) in the framework of SFB-925 – project 170620586 - and the excellence cluster 'Advanced Imaging of Matter' - EXC 2056 - project 390715994.
%%%
%%%
%%%
%

%%%
%%%
%%%
\clearpage
\setcounter{figure}{0}
\renewcommand{\figurename}{Supplementary Fig.}
\renewcommand{\tablename}{Supplementary Table}
\renewcommand{\theequation}{S\arabic{equation}}
\renewcommand{\thefigure}{S\arabic{figure}}
\renewcommand{\thetable}{S\arabic{table}}
\section*{Supplementary Materials}
\subsection*{Sample preparation and experimental procedure}
For the preparation of our homogeneous 2D Fermi gases, we use the experimental setup and procedure described in \cite{Hueck2018-sx,Sobirey2021-gk}. 
In brief, we use first laser- and then evaporative cooling to prepare ultracold gases of approximately $6000$ fermionic $^6$Li atoms in a balanced mixture of the two lowest-energy hyperfine states.
In the radial direction, the gas is held in place by a repulsive ring potential with a diameter of about $70\,\mu m$, resulting in a density per spin state of $n=0.8\,\mathrm{atoms}/\mu m^{2}$, corresponding to $E_{\rm F} \approx h \cdot 8.5\,\mathrm{kHz}$. 
In the vertical direction, an optical lattice provides a tight harmonic confinement with a trapping frequency of $\hbar \omega_z = h \cdot 9.2\,\mathrm{kHz}$.
To control the interparticle interactions, we apply magnetic offset fields between 700\,G and 1000\,G.
Due to the presence of a Feshbach resonance at a magnetic field of 832\,G \cite{Zurn2013-sk}, this allows us to access interaction strengths throughout the BEC-BCS crossover.
By ensuring that all interaction ramps are adiabatic, we keep the system at an approximately constant entropy per particle for all experiments.
The entropy in our system corresponds to a temperature of $T=0.04(1)\,T_{\rm F}$ at an interaction strength of $\rm{ln}(k_{\rm F}a_{\rm 2D}) = -2.9$, which according to \cite{Sobirey2021-gk} is well below the critical entropy for superfluidity in the crossover regime.

\subsection*{Influence of the third dimension}
To obtain a quasi-2D system for our experiments, we trap our atoms in a highly anisotropic confinement, where the spacing of the energy levels in the tightly confined direction is much larger than the reduced chemical potential $\tilde{\mu}=\mu+E_{\rm B}/2$ and the temperature $T$ of the gas.
In this regime, the interactions between the particles can be mapped onto an effective 2D interaction, and the long-range physics of the system become essentially two-dimensional \cite{Petrov2001-qx}.
However, on length scales that are comparable to the size of the tight confinement, there is still a residual influence from the third dimension that causes these short-range physics to be different from the ones found in a purely 2D system.
Performing a quantitative comparison with purely 2D theories therefore requires taking into account these corrections.
The most important difference compared to a purely 2D system is a modification of the two-body binding energy as discussed in \cite{Levinsen2015-lf}, which for a quasi-2D geometry is given by
\begin{equation}
\label{bindingenergy}
\frac{l_{\rm z}}{a_{\rm 3D}} = \int_0^\infty \frac{du}{\sqrt{4 \pi u^3}} \left( \frac{\exp(\frac{-E_{\rm B}}{\hbar \omega_z} u)}{\sqrt{\frac{1}{2 u}(1-\exp(-2 u))}} \right)\,,
\end{equation}
where $l_{\rm z}$ is the harmonic oscillator length of the tight harmonic confinement \cite{Levinsen2015-lf}.
Equation~\ref{bindingenergy} has been found to be in good agreement with experiments, see e.g. \cite{Levinsen2015-lf,Murthy2018-en}. 
This modification of the binding energy is important when comparing our data to QMC results in Fig.~\ref{FigComp}, as these are obtained from purely 2D calculations. 
We therefore we use the QMC results for the many-body contribution $\Delta-E_{\rm B}/2$ and eq.~\ref{bindingenergy} for the binding energy.

The influence of the third dimension also becomes relevant when there is a significant population of higher energy levels in the strongly confined direction. 
In our systems, the temperature is negligible compared to the level spacing ($T<0.1\,\hbar \omega_z$) \cite{Sobirey2021-gk}, and we therefore neglect the influence of thermal excitations in the strongly confined direction. 
The second relevant quantity is the reduced chemical potential $\tilde{\mu}$.
While $\tilde{\mu}$ increases towards the BCS regime \cite{Shi2015-pd}, it remains well below the level spacing $\hbar \omega_z$ for our experiments.
Nevertheless, for our larger values of $\tilde{\mu}/\hbar \omega_z$ there are modifications of the scattering physics of the system, which are taken into account in the definition of $a_{\rm 2D}$ \cite{Petrov2001-qx,Turlapov2017-wg}.

\subsection*{Measuring the dynamic structure factor.}
We obtain $S(q,\omega)$ by moving a weak Bragg lattice through the gas and measuring the resulting heating rate.
The Bragg lattice is formed by two far-detuned laser beams focused onto the atoms with a high-resolution objective as described in \cite{Sobirey2021-gk}.
Two acousto-optic modulators set the frequency difference of the two beams, while two motorized translation stages can be used to change their crossing angle.
This enables us to perform two-photon spectroscopy of the gas with tunable energy and momentum transfer.
To determine the heating rate caused by the Bragg lattice, we relate the decrease in the condensate fraction after an adiabatic interaction ramp to the BEC regime to the increase in the energy $E$ of the system \cite{Sobirey2021-gk}.
This in turn gives us access to the dynamic structure factor via the relation $S(q,\omega) \propto \frac{1}{\omega}\frac{dE}{dt}$ \cite{Brunello2001-hk}. 
While thermal occupation of excitations can modify the measured spectrum \cite{Kuhnle2011-sd}, these effects are negligible for the low temperatures and comparatively high energy transfers in our experiments.
We note that the increased noise in the measured dynamic structure factors at low frequencies is an artifact of the division by $\omega$ and does not suggest the presence of actual excitations in the system.

\subsection*{Chemical potential and binding energy}
To plot our data as a function of the chemical potential in Fig.~\ref{FigComp} and Fig.~\ref{FigMat}\textbf{a}, we need to perform a conversion between the 2D and 3D interaction parameters and the chemical potential. 
For the 3D data, we use the results of fixed-node diffusion Monte Carlo calculations performed in \cite{Astrakharchik2004-qs}, which have been shown to be in good agreement with measurements of the equation of state in 3D Fermi gases \cite{Navon2010-bz,Ku2012-wd}.
For the 2D system, we use the auxiliary-field QMC calculations performed in \cite{Shi2015-pd}, which are in good agreement with recent measurements \cite{Boettcher2016-hu,Fenech2016-ei,Bohlen2020-iu}.
We can therefore determine the chemical potential for our quasi-2D system by taking these results and including the appropriate binding energy given by eq.~\ref{bindingenergy}. 

\subsection*{Determination of the superfluid gap.}
To determine the size of the superfluid gap from our measurements of the dynamic structure factor, we employ two different methods.
The first is based on taking measurements of $S(q,\omega)$ as shown in  Fig.\,\ref{FigIntro} and integrating over the momentum axis to obtain $S(\omega)$.
While the exact line shape of $S(\omega)$ is unknown, we expect it to be qualitatively similar to the one observed in the Raman response of s-wave BCS superconductors, which exhibits a sharp onset at an energy of $2 \Delta$ followed by a slow decay.
However, the presence of the Goldstone mode in neutral superfluids and additional effects such as Fourier broadening and density inhomogeneities modify the line shape and lead to nonzero values of $S(\omega)$ below $2\Delta$.
We therefore fit our measurements of $S(\omega)$ with a phenomenological line shape consisting of a linear increase followed by a Gaussian decay, where the intersection point of the two parts is identified as $2 \Delta$.
The second method is based on the work performed in \cite{Hoinka2017-ej}.
It relies on measuring $S(q,\omega)$ at a fixed momentum transfer $q$ that is chosen such that the pair breaking mode is well separated from the low-energy Goldstone mode. 
This allows us to perform a simple bilinear fit to determine the onset of the pair breaking feature (see Fig.~\ref{FigGap}\textbf{b}), with the intersection point identified as $2 \Delta$.
Note that for large regions of the interaction strength we employ both methods to measure the superfluid gap and find very similar results.

\subsection*{Pair size determination}
As stated in the main text, the pair size $\xi$ can serve as an interaction parameter that is independent of the dimensionality of the system \cite{Pistolesi1994-oe,Marsiglio2015-qh} and is known for a large number of solid-state superfluids (see Table \ref{ExTableMaterials}).
However, for ultracold gases there is only a single study of the pair size in a 3D Fermi gas \cite{Schunck2008-us}, and no pair size measurements have been performed in 2D Fermi gases at all.
Therefore, we make use of the momentum dependence of pair breaking excitations to obtain an estimate for the pair size $\xi$ from our measurements of the dynamic structure factor $S(q,\omega)$.
In particular, we consider the suppression of pair breaking excitations that occurs when the wavelength $\lambda_0 = 2\pi/q$ of the probing lattice becomes comparable to the size of the pairs.
This suppression can be understood by considering the differential force that acts on the constituents of a pair: If the wavelength of the Bragg lattice is much larger than the pair size, the differential force vanishes and pair breaking becomes strongly suppressed. 
Consequently, the onset of the pair breaking continuum in the momentum axis is directly related to the size of the fermion pairs.

\begin{figure*}
\begin{center}
\includegraphics{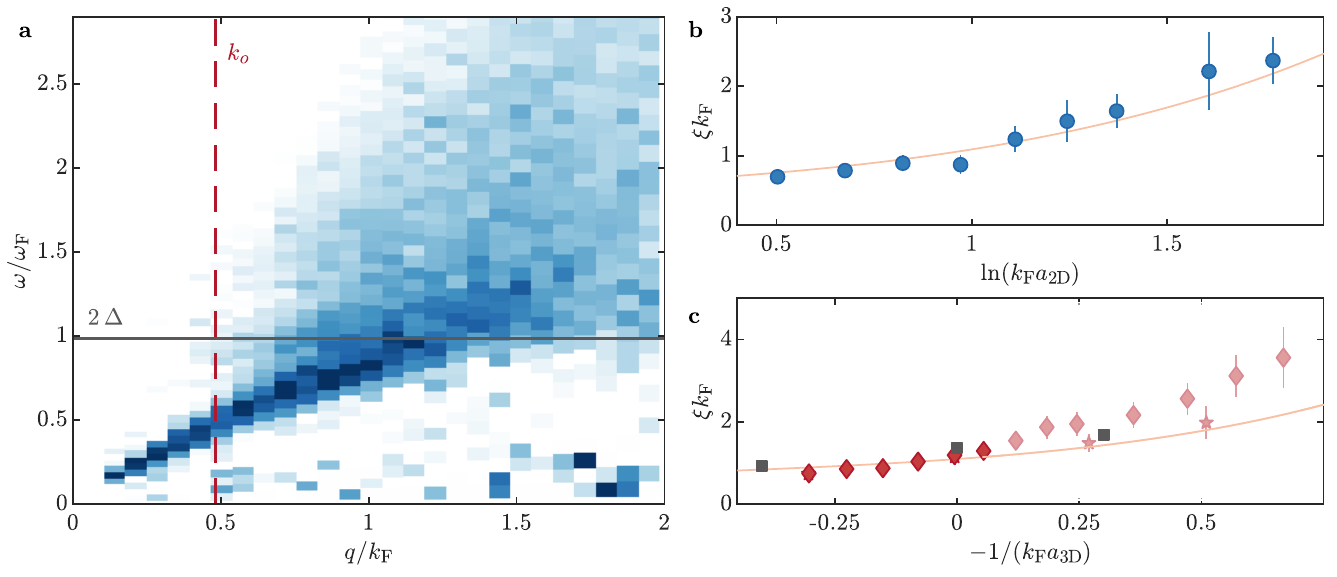}
\caption{\textbf{Pair size in the BEC-BCS crossover} 
\textbf{a}: Estimation of the pair size from a measurement of the dynamic structure factor.
We determine the onset momentum $k_o$ (red-dashed vertical line) of the pair breaking continuum via a bilinear fit of the dynamic structure factor $S(q,2 \Delta)$ at a transferred energy of $\hbar \omega = 2 \Delta$ (black horizontal line).
We then obtain an estimate of the pair size $\xi \approx \alpha_{3D}/k_o$.
\textbf{b}: Estimate of the pair size $\xi = \alpha_{2D}/k_o$ for a 2D Fermi gas (blue circles).
The value of $\alpha_{2D} \approx 0.6$ is obtained by matching our results for $k_o$ to the theoretical prediction \cite{Randeria1990-ox} (solid line).
\textbf{c}: Pair size $\xi = \alpha_{3D}/k_o$ for a 3D Fermi gas.
Red stars denote data points where $k_o$ was extracted from full measurements of the 3D structure factor as shown in \textbf{a}, red diamonds denote data from a separate set of measurements where data was only taken at transferred energies of $\hbar \omega \approx 2 \Delta$.
The solid line denotes a theory prediction from ref.~\cite{Marini1998-bb}, gray squares show results from a pair size measurement performed in ref.~\cite{Schunck2008-us}, rescaled as discussed in ref.~\cite{Palestini2014-jw}.
Error bars denote $1\sigma$ confidence intervals of the fit.
}
\label{Fig_SM_Pairsize}
\end{center}
\end{figure*}

For our estimate of the pair size, we make the simple approximation that we can define an onset momentum $k_o$ that is related to the pair size via $k_o = \alpha/\xi$ with a proportionality factor $\alpha$.
We identify $k_o$ as the intersection point of a bilinear fit which is performed on a slice through the dynamic structure factor at a transferred energy of $\hbar \omega = 2 \Delta$ (see Fig.~\ref{Fig_SM_Pairsize}\textbf{a}).
To estimate the value of $\alpha$, we compare our results for $1/k_o$ to theoretical predictions of $\xi$ for the given dimensionality.

Fig.~\ref{Fig_SM_Pairsize}\textbf{b} shows the theoretical prediction for the pair size of a 2D Fermi gas from mean-field theory \cite{Randeria1990-ox} together with our estimate of $\xi$, with a prefactor of $\alpha_{2D} \approx 0.6$. 
We find that the interaction dependence of our estimate is in excellent agreement with the mean-field prediction. 

The comparison between the prediction for the pair size of a 3D Fermi gas from ref.~\cite{Marini1998-bb} and our estimate of $\xi$ using a prefactor of $\alpha_{3D} \approx 0.8$ is shown in Fig.~\ref{Fig_SM_Pairsize}\textbf{c}.
As the determination of $k_o$ becomes increasingly difficult towards the BCS regime where $k_o$ approaches the smallest transferred momenta we can realize in our experimental setup, only measurements for $1/k_{\rm F}a_{\rm 3D} > -0.1$ were considered in the determination of $\alpha$.
In this interaction range, the interaction dependence of our estimate of $\xi$ is then in good agreement with the theoretical prediction.

In addition, our estimate of $\xi$ in 3D is in excellent agreement with the measurements of the pair size of a 3D Fermi gas performed in \cite{Schunck2008-us} using RF spectroscopy (black squares in Fig.~\ref{Fig_SM_Pairsize}\textbf{c}). 
This excellent agreement, as well as the very similar values of $\alpha$ for the 2D and 3D systems, strongly support the validity of our approach to determine $\xi$ from our measurements of the dynamic structure factor.
Additionally, we note that small changes in our estimate of $\xi$ do not significantly affect the conclusions drawn from Fig.~\ref{FigMat}\textbf{b}.

\begin{table*}
\begin{center}
\begin{tabularx}{10cm}{X l l l l r}
\hline
Material & $\nicefrac{\Delta}{k_B}$ & $\nicefrac{E_{\rm F}}{k_B}$ & $\xi$ & $k_{\rm F}$ & Ref. \\
 & [K] & [K] & [\r{A}] & [\r{A}$^{-1}$] & \\
\hline
Bi$_2$Sr$_2$CaCu$_2$O$_8$ & 140 & 970 & 7.0 & 0.41 & \cite{Harshman1992-jg,Yoshida2009-qo}\\
Tl$_2$Ba$_2$Ca$_2$Cu$_3$O$_{10}$ & 270 & 1800 & 7.3 & 0.49 & \cite{Harshman1992-jg,Chia2011-jx,Ponomarev2014-mv}\\ 
La$_{1.85}$Sr$_{0.15}$CuO$_4$ & 84 & 1100 & 11 & 0.46 & \cite{Harshman1992-jg,Yoshida2009-qo}\\
YBa$_2$Cu$_3$O$_7$ & 470 & 8800 & 15 & 0.45 & \cite{Harshman1992-jg,Grissonnanche2014-nn,Dagan2000-kh} \\ 
RbCa$_2$Fe$_4$As$_4$F$_2$ & 95 & 730 & 10 & 0.20 & \cite{Adroja2018-mi} \\
FeSe/STO & 240 & 6200 & 6.7 & 0.61 &  \cite{Ge2015-es,Biswas2018-eo}\\
Pb & 13 & 120000 & 830 & 1.6 & \cite{Carter1995-up} \\
Sn & 6.6 & 110000 & 2300 & 1.6 & \cite{Carter1995-up} \\
Al & 2.1 & 110000 & 16000 & 1.6 & \cite{Carter1995-up} \\
Nb & 18 & 13000 & 111 & 0.54 & \cite{Richards1960-df,Townsend1962-vz,Nakagawa2021-ih} \\
BaKBiO$_3$ & 61 & 2900 & 53 & 0.26 & \cite{Escudero1994-ov,Carter1995-up} \\ 
$\kappa$-(ET)$_2$Cu(NCS)$_2$ & 25 & 210 & 23 & 0.17 & \cite{Harshman1992-jg,Wosnitza2003-gg} \\
NbSe$_2$ & 15 & 810 & 80 & 0.54 & \cite{Le1991-gq,Dvir2018-gx}\\
PbMo$_6$S$_8$ & 36 & 1600 & 9.8 & 0.31 & \cite{Harshman1992-jg,Petrovic2011-jj}\\
YB$_6$ & 16 & 2400 & 330 & 0.41 & \cite{Hillier1997-xk}\\
YNi$_2$B$_2$C & 21 & 4200 & 80 & 0.94 & \cite{Hillier1997-xk}\\
LuRuB$_2$ & 16 & 4100 & 180 & 0.85 & \cite{Lee1987-za,Barker2018-cv}\\
YRuB$_2$ & 13 & 2400 & 110 & 0.84 & \cite{Lee1987-za,Barker2018-cv}\\
K$_3$C$_{60}$ & 33 & 470 & 26 & 0.34 & \cite{Uemura1991-nv,Degiorgi1994-jb}\\
Rb$_3$C$_{60}$ & 62 & 2900 & 29 & 0.49 & \cite{Degiorgi1994-jb,Carter1995-up,Varshney2012-pw}\\
Re$_3$Ta & 8.6 & 640 & 110 & 0.46 & \cite{Barker2018-io} \\
Re$_{5.5}$Ta & 16 & 2000 & 45 & 0.44 & \cite{Singh2020-aw}\\
Nb$_{0.5}$Os$_{0.5}$ & 5.6 & 660 & 78 & 0.15 & \cite{Singh2018-ob} \\
La$_7$Ir$_3$ & 4.3 & 520 & 190 & 0.40 & \cite{Barker2015-fo}\\
LaPtGe & 4.9 & 1200 & 59 & 0.43 & \cite{Sajilesh2018-bq}\\
BeAu & 5.8 & 18000 & 1900 & 1.1 & \cite{Singh2019-td}\\
TaOs & 3.5 & 900 & 74 & 0.50 & \cite{Singh2017-kj}\\ 
Zr$_3$Ir & 4.1 & 1700 & 46 & 0.75 & \cite{Singh2019-ri}\\
NbOs$_2$ & 4.8 & 620 & 90 & 0.43 & \cite{Singh2019-ag}\\ 
$^3$He & 0.0046 & 0.89 & 200 & 0.78 &  \cite{Movshovich1990-ou,Pistolesi1994-oe}\\ 
Li$_x$ZrNCl & \multicolumn{4}{c}{tuneable} & \cite{Nakagawa2021-ih} \\
\hline
\end{tabularx}
\caption{\textbf{Parameters for different fermionic superfluids in Fig.\,\ref{FigMat}.}
Values and references for the superconducting gap $\Delta$, the Fermi energy $E_{\rm F}$, the coherence length $\xi$ and the Fermi wavevector $k_{\rm F}$ used in Fig.~\ref{FigMat}\textbf{b}.
For materials such as cuprate superconductors where these parameters are dependent on the doping level, results for optimally doped samples were used. 
We note that there are different extrapolation methods and definitions of the coherence length, leading to different values between references.
However, the differences are generally on the order of unity and therefore do not affect the conclusions drawn from Fig.~\ref{FigMat}\textbf{b}.
}
\label{ExTableMaterials}
\end{center}
\end{table*}

\begin{table*}
\begin{center}
\begin{tabular}{ccrrr|ccrrr}
\hline
B [G] & $\ln(k_{\rm F} a_{\rm 2D})$ & $\nicefrac{\Delta}{E_{\rm F}}$ & $\nicefrac{\mu}{E_{\rm F}}$ & $\xi k_{\rm F}$ & B [G] & $(k_{\rm F} a_{\rm 3D})^{-1}$ & $\nicefrac{\Delta}{E_{\rm F}}$ & $\nicefrac{\mu}{E_{\rm F}}$ & $\xi k_{\rm F}$\\
\hline
\multicolumn{4}{c}{\textbf{2D}, Low $q$}&&&\multicolumn{4}{c}{\textbf{3D}, Low $q$}   \\ \cline{1-5} \cline{6-10}
784 & -0.55 & 1.084(4) & -0.83 &    - & 812 &  0.37 & 0.64(7) & 0.08 & 0.73 \\
792 & -0.29 & 0.98(1) & -0.53 &    - & 816 &  0.27 & 0.573(7) & 0.19 & 0.82 \\
800 & -0.07 & 0.87(1) & -0.31 &    - & 820 &  0.22 & 0.58(1) & 0.27 & 0.89 \\
808 &  0.16 & 0.77(1) & -0.11 &    - & 824 &  0.15 & 0.48(2) & 0.32 & 1.01 \\
816 &  0.33 & 0.71(1) &  0.01 & 0.70 & 828 &  0.08 & 0.497(7) & 0.38 & 1.18 \\
824 &  0.50 & 0.64(1) &  0.12 & 0.76 & 832 &  0.01 & 0.47(1) & 0.44 & 1.41 \\
855 &  1.10 & 0.47(2) &  0.39 & 1.23 & 835 & -0.05 & 0.44(2) & 0.48 & 1.63 \\
871 &  1.34 & 0.423(6) &  0.46 & 1.52 & 839 & -0.12 & 0.384(5) & 0.53 & 1.93 \\
886 &  1.58 & 0.410(5) &  0.52 & 1.86 & 843 & -0.18 & 0.38(1) & 0.56 & 2.22 \\
    &  		&		  &		  &      & 847 & -0.24 & 0.34(1) & 0.59 & 2.54 \\
\multicolumn{4}{c}{\textbf{2D}, $S(\omega)$}&&855&-0.35&0.26(1)&0.64& 3.16 \\\cline{1-5}
824 & 0.50  & 0.61(1) & 0.13  & 0.76 &   &       &         &      &      \\
832 & 0.68  & 0.526(9) & 0.22  & 0.85 &&\multicolumn{4}{c}{\textbf{3D}, High $q$}\\\cline{6-10}
839 & 0.82  & 0.55(1) & 0.29  & 0.96 & 863 & -0.43 & 0.222(3) & 0.67 & 3.63 \\
847 & 0.97  & 0.50(1) & 0.35  & 1.09 & 867 & -0.51 & 0.24(2) & 0.69 & 4.04 \\
855 & 1.11  & 0.493(9) & 0.40  & 1.24 & 871 & -0.53 & 0.193(7) & 0.69 & 4.11 \\
863 & 1.24  & 0.42(1) & 0.44  & 1.40 & 879 & -0.62 & 0.1686(5) & 0.71 & 4.51 \\
871 & 1.37  & 0.46(1) & 0.47  & 1.56 & 886 & -0.73 & 0.16(1) & 0.73 & -    \\ 
886 & 1.61  & 0.383(5) & 0.52  & 1.91 & 886 & -0.74 & 0.14(2) & 0.73 & -    \\
898 & 1.77  & 0.360(8) & 0.55  & 2.17 & 894 & -0.83 & 0.126(2) & 0.74 & -    \\
910 & 1.92  & 0.31(1) & 0.58  & 2.42 & 902 & -0.91 & 0.11(1) & 0.75 & -    \\
926 & 2.11  & 0.25(2) & 0.61  & 2.75 & 910 & -0.98 & 0.057(7) & 0.76 & -    \\
\hline
\end{tabular}
\caption{\textbf{Summary of experimental results.}
Shown are the measured values of the gap for both 2D and 3D Fermi gases, with the values for the 3D Fermi gases taken from \cite{Biss2021-ir}.
Also shown are the values of the chemical potential $\mu$ used in Fig.~\ref{FigComp} (see Supplementary Materials) and the approximate value of $\xi k_{\rm F}$ obtained from a spline fit to the data shown in Fig.~\ref{FigMat}\textbf{a}.
Errors denote $1\sigma$ confidence intervals of the fit.
}
\label{ExTableData}
\end{center}
\end{table*}

\begin{figure*}
\begin{center}
\includegraphics[width=18.3cm]{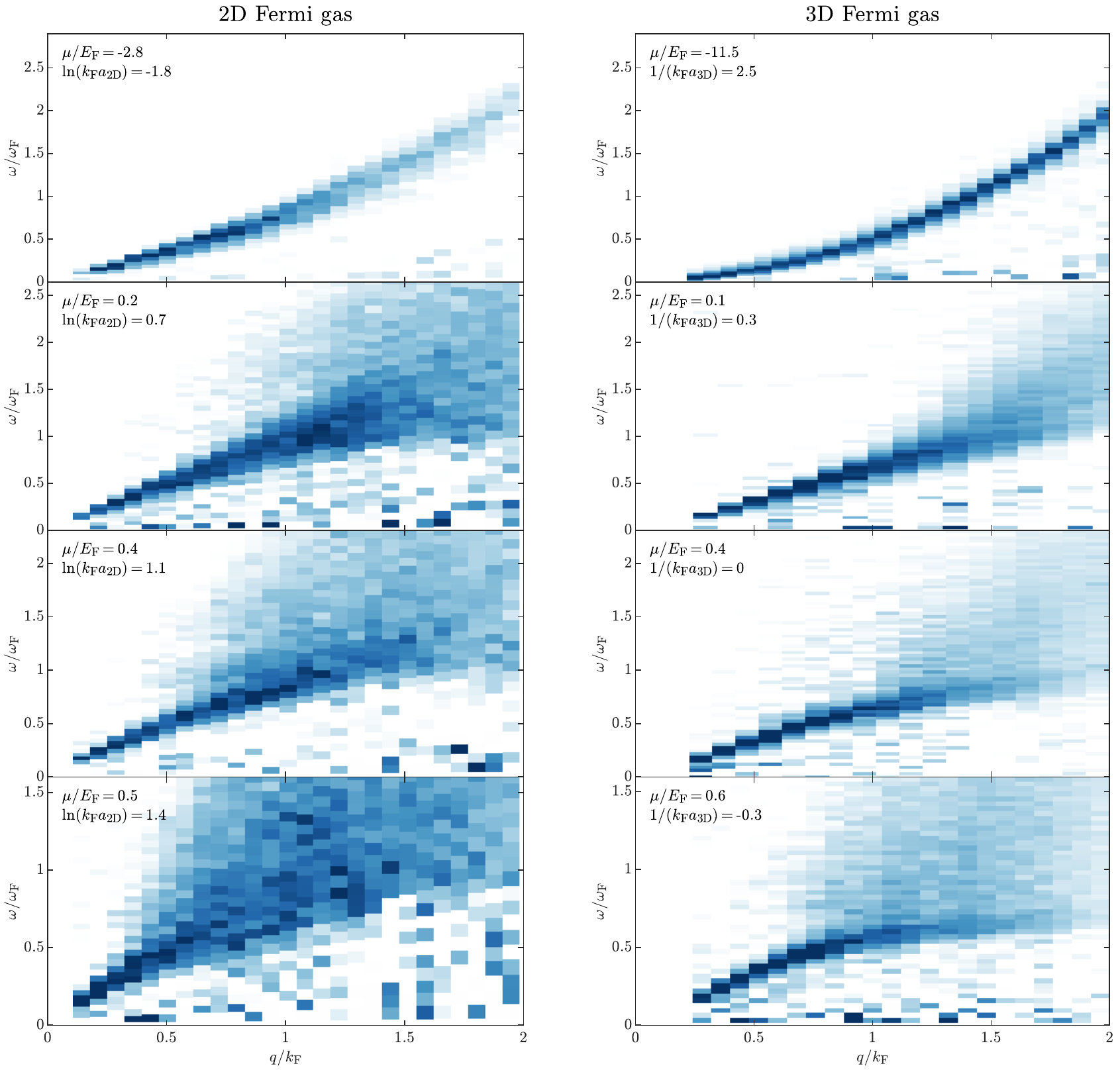}
\caption{\textbf{Measurements of the dynamic structure factor in two- and three-dimensional ultracold Fermi gases.} 
The excitation spectra of systems with comparable values of $\mu/E_{\rm F}$ show very similar qualitative behavior.
3D data taken from \cite{Biss2021-ir}.
}
\label{Fig_SM_Comparison}
\end{center}
\end{figure*}

\clearpage
\pagebreak


\begin{thebibliography}{53}
\bibitem [{Not()}]{NoteMu}%
  \BibitemOpen
  \href@noop {} {}\bibinfo {note} {Note that we plot the gap from
  \cite{Haussmann2007-ph} against the chemical potential from
  \cite{Astrakharchik2004-qs} instead of the value for $\mu/E_{\rm F}$ obtained
  in \cite{Haussmann2007-ph}. However, as the two results for the chemical
  potential are very similar, this does not significantly affect the comparison
  or its conclusion.}\BibitemShut {Stop}%
\end{thebibliography}
\end{document}